\documentclass{aa}  
\usepackage[normalem]{ulem}
\usepackage{placeins}
\usepackage{color}
\usepackage{calc}
\usepackage{amsmath,amssymb,graphicx}
\usepackage{tensor}
\usepackage{bm}
\usepackage{times}
\usepackage[varg]{txfonts}
\usepackage{float}
\usepackage{dcolumn}
\usepackage[nolist,nohyperlinks]{acronym}
\usepackage{xspace}
\usepackage[abs]{overpic}
\usepackage{pict2e}
\usepackage{enumitem}
\usepackage[usenames,dvipsnames]{xcolor}
\usepackage[utf8]{inputenc}
\usepackage{acronym}
\usepackage{gensymb}

\usepackage[normalem]{ulem}
\usepackage{longtable}
\usepackage{verbatim}
\usepackage{multirow}
\usepackage{amsmath,amssymb}
\usepackage{csquotes}
\usepackage[export]{adjustbox}

\usepackage{subfigure}
\usepackage{placeins}
\usepackage[pdfborder={0 0 0},colorlinks=true,citecolor=NavyBlue]{hyperref}
\usepackage{linenoaa}
\usepackage{soul}

\usepackage{cleveref}

\begin{document}

\title{The interacting double white dwarf population with LISA: Stochastic foreground and resolved sources}

\author{A. Toubiana \inst{1} \and N. Karnesis \inst{2} \and A. Lamberts \inst{3,4}  \and M. C. Miller \inst{5}}

\institute{Max Planck Institute for Gravitationsphysik (Albert Einstein Institute), Am M\"{u}hlenberg 1, 14476 Potsdam, Germany  \\
\and
Department of Physics, Aristotle University of Thessaloniki, Thessaloniki 54124, Greece  \\
\and
  Laboratoire Lagrange, Université Côte d’Azur, Observatoire de la Côte d’Azur, CNRS, Bd de l’Observatoire, 06300, France \\
\and 
Laboratoire Artemis, Université Côte d’Azur, Observatoire de la Côte d’Azur, CNRS, Bd de l’Observatoire, 06300, France \\
\and 
Department of Astronomy and Joint Space-Science Institute, University of Maryland, College Park, MD 20742-2421, USA
}
 
  \abstract
   {}
   {We investigate the impact of tidal torques and mass transfer on the population of double white dwarfs that will be observed with LISA.}
  {Our Galactic distribution of double white dwarfs is based on the combination of a cosmological simulation and a binary population synthesis model. We used a semi-analytical model to evolve double white dwarf binaries considering ten different hypotheses for the efficiency of tidal coupling and three hypotheses for the birth spins of white dwarfs. We then estimated the stochastic foreground and the population of resolvable binaries for LISA for these different combinations.}
   {Our predicted double white dwarf binary distribution can differ substantially from the distribution expected if only gravitational waves (GWs) are considered. If white dwarfs spin slowly, then we predict an excess of systems around a few mHz, due to binaries that outspiral after the onset of mass transfer. This excess of systems leads to differences in the confusion noise, which are most pronounced for strong tidal coupling. In that case, we find a significantly higher number of resolvable binaries than in the GW-only scenario. If instead white dwarfs spin rapidly and tidal coupling is weak, then we find no excess around a few mHz, and the confusion noise due to double white dwarfs is very low. In that scenario, we also predict a subpopulation of outspiralling binaries below 0.1 mHz. Using the Fisher matrix approximation, we estimate the uncertainty on the GW-frequency derivative of resolvable systems. We find that, even for non-accreting systems, the mismodelling error due to neglecting effects other than GWs is larger than the statistical uncertainty, and thus this neglect would lead to biased estimates for mass and distance.} 
   {Our results suggest that the population of double white dwarfs is likely to be different from what is expected in the standard picture, which highlights the need for flexible tools in LISA data analysis. Because our semi-analytical model hinges upon a simplistic approach to determining the stability of mass accretion, it will be important to deepen our comprehension of stability in mass-transferring double white dwarf binaries.}


   \maketitle

\section{Introduction}

The Milky Way contains $\sim 10^7$ double white dwarf (DWD) binaries, which  emit nearly monochromatic gravitational waves (GWs) within the mHz band \citep{Breivik:2017jip,Nelemans:2001hp,Ruiter:2007xx,Lamberts:2019nyk}. The Laser Interferometer Space Antenna (LISA)~\citep{LISA:2017pwj}, which is scheduled for launch in 2035, will complement electromagnetic instruments with gravitational wave (GW) detection, and will thus provide a unique opportunity to study this largely unobserved population. Unresolvable DWDs are expected to produce stochastic noise that will exceed LISA's instrumental noise in the $\sim 0.1-2$~mHz range. Additionally, up to tens of thousands of binaries are expected to be individually resolved. This makes the analysis of LISA data challenging.  This challenge has spurred the development of {\it Global Fit} pipelines~\citep{Littenberg:2023xpl}, which are designed to infer this persistent yet unknown signal present in the data. The detection and characterisation of this population will help us understand the formation and evolution of white dwarfs (WDs), which is one of the eight objectives of the LISA mission \citep{Colpi:2024xhw}. Now that LISA has been adopted by the European Space Agency, it is timely to construct the tools for the analysis of LISA data and its astrophysical interpretation. Similarly to the recent study of \citet{Scaringi:2023xpm}, which highlighted observational signatures of cataclysmic variables within LISA's confusion noise, our work anticipates distinct observational features, attributable in our case to tidal effects in mass-transferring binaries. 

The evolution of DWDs is characterised by various physical processes that can significantly influence their behaviour and observational signatures. One of the key aspects is the effect of matter interactions, which becomes relevant early in the evolution, due to the reduced compactness of WDs compared to neutron stars and black holes. 
Similar to binary stars, tidal torques work to synchronise the rotation period of the WDs with the orbital period of the binary~\citep{1977A&A....57..383Z,Verbunt:1983ar}. While expressions for the intensity of these torques within DWDs have been proposed~\citep{10.1093/mnras/207.3.433}, they are generally valid only under the assumption of low asynchronicity, and the precise intensity of these torques remains uncertain.
As the separation between the WDs decreases, the lighter WD may overflow its Roche lobe, leading to mass transfer onto the heavier WD. This scenario can result in one of two outcomes. If the mass transfer is dynamically unstable, the binary may merge rapidly, potentially triggering a Type Ia supernova~\citep{1984ApJ...277..355W,1986ARA&A..24..205W,2007A&A...476.1133F,2010A&A...514A..53F,2010ApJ...709L..64G,2010ApJ...719.1067K,2014ARA&A..52..107M,2018ApJ...854...52S}. If instead the mass transfer is stable, then an AM CVn system can be formed \citep{1981ApJ...244..269N,2001A&A...365..491N,2010PASP..122.1133S,2002ARep...46..667T}. These stable interacting binaries offer unique opportunities for multi-messenger observations~\citep{Nelemans:2003ha,Shah:2014nea,Kupfer:2019ngt,Kupfer:2023nqx}. 

\cite{Marsh:2003rd} proposed the first semi-analytical model for the evolution of DWDs including tidal torques and mass transfer, in addition to GWs. This approach was extended in~\cite{Gokhale:2006yn} to include the evolution of the lighter WD spin. Finally, \cite{2015ApJ...806...76K} proposed a variation of the previous models with a more refined treatment of mass transfer based on the ballistic accretion of particles~\citep{Sepinsky:2014ila}. Those works highlighted the effect of tidal torques on the evolution of DWDs, and in particular that strong tides can help stabilise the dynamics of mass transfer, preventing rapid mergers. We note that in these analyses, the stability of mass transfer is determined using a simple and thus potentially unrealistic criterion based on the rate of mass lost by the donor. 

Systems surviving mass transfer are predicted to outspiral (i.e. they will have a negative orbital frequency derivative). This distinctive signature can be used to identify the subpopulation of mass-transferring DWDs with LISA, as was proposed in~\cite{Breivik:2017jip}, \cite{Kremer:2017xrg}, \cite{Biscoveanu:2022sul} and \cite{Yi:2023osk}. In particular, in \cite{Biscoveanu:2022sul}, the authors simulate a population of DWDs until the onset of mass transfer using the \texttt{COSMIC} population synthesis code \citep{Breivik:2019lmt} and use the semi-analytical model to compute the long-term evolution of DWDs. They explore three distinct assumptions regarding the intensity of tidal coupling, and perform a population analysis on these DWDs to demonstrate the potential of LISA observations to probe the intensity of tidal coupling within these systems. They also investigate the impact of different assumptions on the common envelope phase during the evolution of DWDs progenitors, complementing the work of~\cite{Kremer:2017xrg}.     
Those works shed light on the potential of LISA to inform us about the outcomes of stellar evolution and about the properties of DWDs. However, the tools currently used to perform population analyses~\citep{Mandel:2018mve} are not applicable to DWDs because the signals overlap and, therefore, are not statistically independent. Moreover, it is necessary to account for the dependence of the contribution of DWDs to the confusion noise on the properties of the population that we seek to infer. 

Alternatively, approaches to measure mass transfer effects in individual binaries have been developed~\citep{Breivik:2017jip,Yi:2023osk}. For non-accreting systems, it has been suggested that it would be possible to measure the chirp mass, and distinguish between WDs and black holes in Galactic binaries~\citep{Sesana:2019jmu,Sberna:2020ycl}, to measure tidal effects~\citep{Shah:2015cua}, and possibly even to perform tests of general relativity (GR)~\citep{Littenberg:2018xxx}.


In our study, we investigate for the first time the impact of tidal effects within DWDs on the confusion noise that they produce and on the properties of resolvable sources. Starting from a population of DWDs at formation predicted by the population synthesis code of ~\cite{Lamberts:2019nyk}, we use a semi-analytical model, detailed in Sec.~\ref{sec:formalism}, similar to those of~\cite{Marsh:2003rd} and \cite{Sberna:2020ycl} to evolve the population until today. We discuss the main features of the long-term evolution of accreting systems in Sec.~\ref{sec:exs}. In this semi-analytical model, tidal torques depend on  (i) the spin of the WDs and (ii) the tidal synchronization timescale. The latter measures the efficiency of tidal coupling: the shorter the synchronization timescale, the stronger the tidal interaction. We introduce a universal parametrisation for the tidal synchronization timescale, which allows the intensity of tidal torques to depend on a single parameter for all DWDs (apart from the spin). We generate mock populations exploring a total of 30 combinations of initial spin and tidal synchronization timescale in Sec.~\ref{sec:pop}, and in Sec.~\ref{sec:da}, we estimate the confusion noise due to DWDs along with the number of detectable binaries, using the methodology of~\cite{Karnesis:2021tsh}, \cite{tim06}, \cite{cro07}, \cite{nis12} and \cite{Lamberts:2019nyk}.

We find significant impacts on the DWD population. In particular, if WDs are born slowly spinning and if tidal effects are strong, we anticipate a detectable excess of events around 1~mHz due to accreting binaries that outspiral and accumulate around that frequency. This
excess affects both the confusion noise and the resolvable binaries. The latter is in agreement with the findings of~\cite{Biscoveanu:2022sul}. Moreover, we estimate how many resolvable systems have measurable GW-frequency derivative. We find astrophysical effects in the evolution (i.e. other than GW radiation) to be relevant for all systems with measurable GW-frequency derivative, even for non-accreting systems. Thus, measuring properties of the binary without accounting for effects other than GW radiation will lead to biased estimates. Finally, in Sec.~\ref{sec:syst}, we assess the robustness of our conclusions against systematic effects by considering modifications to our semi-analytical model. We find our results to be generally robust, with the main uncertainty lying in the validity of the criterion we employ to determine the stability of mass-transferring DWDs. We present our concluding remarks in Sec.~\ref{sec:ccl} and give some extra details on oscillatory behaviours in the evolution of mass-transferring DWDs in Appendix~\ref{app:res}.

\section{Evolution of double white dwarf binaries}\label{sec:formalism}

Our DWD population is based on the catalogue created by \citet{Lamberts:2019nyk}. This model uses a zoom-in simulation of a galaxy with properties similar to the Milky Way to provide a metallicity-dependent star formation rate that naturally reproduces a thin and thick galactic disk, as well as a bulge and a halo \citep{Wetzel_16_Latte}. This was combined with the output of the binary population synthesis code BSE \citep{Hurley_2002_BSE} with standard parameters. The population was evolved until the formation of the DWD, and we only kept the systems that have reached that stage by the present day. From this point on, we used the semi-analytical described below to evolve the DWDs until today. In comparison,~\citet{Kremer:2017xrg} and \cite{Biscoveanu:2022sul} evolved DWDs until the onset of mass transfer with a population synthesis code and, from there, used a semi-analytical model similar to ours to evolve the binaries for 10 Gyr. Our model naturally produces combinations of WDs with different core compositions (helium, carbon/oxygen or neon), which have different evolutionary timescales and typical masses and sky localisations. In the remainder of this paper, we consider the full population of systems, but do not distinguish the different subtypes of systems. 
In Sec.~\ref{sec:syst}, we discuss the impact of the assumptions we make to model the evolution of DWDs. 
We use $m_{i}$, $\omega_{i}$, $J_{i}$ and $R_{i}$ ($i=1,2$) to represent respectively the masses, spins, angular momenta and radii of the WDs, with $m_1\geq m_2$. We define the mass ratio $q=m_2/m_1 \leq 1$. We denote by $M_t=m_1+m_2$ the total mass of the binary, $a$ its separation, $\omega_o$ its orbital frequency and $J_o$ its orbital angular momentum. We assume all binaries to be quasi-circular. 

We begin by specifying our model for mass transfer.  Using $R_L$ for the Roche lobe radius of the lighter WD, we define the Roche lobe overfill factor as
\begin{equation}
    \Delta=R_2-R_L \label{eq:overfill}.
\end{equation}
We assume that mass is exchanged when the lighter WD overflows its Roche lobe (i.e. when $\Delta>0$). We use Eggleton's approximation \citep{Eggleton:1983rx} for $R_L$, and following~\cite{Marsh:2003rd}, we take
\begin{align}
&\dot{m}_2 =
	\begin{cases}
		& -\mathcal{F}_{\rm acc}(m_2,m_1,a,R_2)\Delta^3 , \  {\rm if} \;  \Delta>0  ,  \nonumber \\ 
		& 0 \ {\rm otherwise} ,
	\end{cases} \label{eq:dm} \\
\end{align} 
where $\mathcal{F}_{\rm acc}$ is defined in Eq. (10) of~\cite{Marsh:2003rd}. We limit the rate of mass accreted by the heavier binary to the Eddington limit or below, using the approximation of~\cite{1999A&A...349L..17H} for the latter: 
\begin{equation}
    \dot{m}_1 = {\rm min(-\dot{m}_2,\dot{m}_{{\rm Edd,1}})}.\label{eq:dm1}
\end{equation}
Therefore, part of the mass might be lost during mass transfer episodes, i.e. we do not necessarily have $\dot{M}_t=0$. 

The evolution of the binary is governed by the angular momentum balance equation. We assume that the system loses angular momentum only via GWs and loss of matter:
\begin{equation}
    \dot{J}_{o}+\dot{J}_1+\dot{J}_2=-\dot{J}_{\rm GW}-\dot{J}_{\rm loss}. \label{eq:j_balance}
\end{equation}
The angular momentum of a WD is related to its spin through \begin{equation}
    J_{i}=k_{i}m_{i}R_{i}\omega_{i}, \label{eq:Ji}
\end{equation}
where $k_{i}$ is a numerical factor such that $k_{i}m_{i}R_{i}^2$ is the moment of inertia of the WD. We use the fit from~\cite{Marsh:2003rd}: \mbox{$k_{i}=0.1939(1.44885-m_{i})^{0.1917}$}. 
We assume that the angular momenta of the WDs evolve due to mass transfer and tidal torques as
\begin{equation}
     \dot{J}_i=j_i\dot{m}_i-\frac{k_im_iR_i^2}{\tau_{s,i}}(\omega_i-\omega_o).  \label{eq:wd_torques}
\end{equation}
In the above equation $\tau_{s,i}$ is the tidal synchronization timescale. The smaller the timescale, the more efficient tidal torques are in synchronising the spin of the WD with the orbital frequency. In~\cite{2020MNRAS.496.5482Y}, the authors highlighted that the linear approximation for tidal torques in Eq.~\eqref{eq:wd_torques} is not expected to hold for the higher frequency DWDs observable by LISA. However, in this work, we adhere to this linear approximation to conduct an initial assessment of the impact of tidal effects on the population of DWDs. While the heavier WD might be spun up through accretion from the companion star that will form the lighter WD, the latter is expected to be slowly rotating at birth, and so its specific angular momentum should contribute little to the evolution of the DWD. Thus, for simplicity, we take $j_2=0$. However, for completeness, as described below, we consider the impact of the donor being rapidly rotating at birth, and, in Sec.~\ref{sec:syst}, we explore another possibility for $j_2$. For the angular momentum of the more massive WD we use the model of~\cite{1988ApJ...332..193V}: 
\begin{equation}
   j_1=\sqrt{Gm_1R_h}. \label{eq:j1}
\end{equation}
Here, $R_h$ is the radius of the orbit of the matter around the accreting WD. A fit for $R_h$ is provided in~\cite{1988ApJ...332..193V}. This formula corresponds to direct impact accretion. After some time, an accretion disk might form. This occurs when the minimum radius reached by the stream, $R_{\rm min}$, exceeds the radius of the accretor. In this case, $R_h$ is replaced by $R_1$ in Eq.~\eqref{eq:j1}. A fit to the minimum radius of the stream is given by~\cite{Nelemans:2001nr}:
\begin{align}
    \frac{R_{\rm min}}{a}=& \  0.04948 \ - \ 0.03815 \log(q) \ + \ 0.04752 \log^2(q)  \nonumber  \\ 
    \ &- \ 0.006973 \log^3(q).
\end{align}
The evolution equation for $\omega_i$ is obtained by taking the derivative of Eq.~\eqref{eq:Ji},
\begin{equation}
    \dot{J}_i=k_im_iR_i^2\omega_i \left ( \lambda_i\frac{\dot{m}_i}{m_i}+\frac{\dot{\omega}_i}{\omega_i} \right ), \label{jdot}
\end{equation}
with $\lambda_i=1+2\frac{{\rm d} \log(R_i)}{{\rm d} \log(m_i)}+\frac{{\rm d} \log(k_i)}{{\rm d} \log(m_i)}$,
and equating it with Eq.~\eqref{eq:wd_torques}, which yields
\begin{equation}
\dot{\omega}_i=\left ( \frac{j_i}{k_iR_i^2}  - \lambda_i \omega_i \right ) \frac{\dot{m}_i}{m_i} - \frac{\omega_i-\omega_o}{\tau_s}. \label{eq:omega}
\end{equation}
Following~\cite{Marsh:2003rd}, we limit the WD spin to the break-up rate of the WD defined as
\begin{equation}
   \omega_{i,K}=\sqrt{\frac{Gm_i}{R_i^3}}. \label{eq:breakup}
\end{equation}
If the WD spin reaches this value, then we enforce $\omega_i=\omega_{i,K}$ and $\dot{\omega}_i=\dot{\omega}_{i,K}$. The latter is achieved by changing the tidal synchronization timescale so that Eq.~\eqref{eq:omega} yields the desired evolution equation for $\omega_i$.

To obtain the evolution equation for the separation, we start from Kepler's law $\omega_o=\frac{\sqrt{Gm}}{a^{3/2}}$, so that the orbital angular momentum can then be written $J_{o}=\sqrt{\frac{Ga}{M_t}}m_1m_2$. Therefore,
\begin{equation}
    \frac{\dot{J}_{o}}{J_{o}}=\frac{1}{2}\frac{\dot{a}}{a}+\frac{\dot{m}_1}{m_1}+\frac{\dot{m}_2}{m_2}-\frac{1}{2}\frac{\dot{M}_t}{M_t}. \label{eq:jorb_dot}
\end{equation}
At Newtonian order: 
\begin{equation}
    \dot{J}_{\rm GW}=\frac{32}{5}\frac{G^3}{c^5}\frac{m_1m_2m}{a^4}J_{o}. \label{eq:j_gw}
\end{equation}
Following~\cite{vanHaaften:2011iy}, we assume an isotropic re-emission of the lost mass and take
\begin{equation}
\dot{J}_{\rm loss}=q\frac{\dot{M}_t}{M_t}J_{o}. \label{eq:jloss}
\end{equation}
Finally, putting together Eqs.~(\ref{eq:j_balance}), (\ref{eq:wd_torques}), (\ref{eq:jorb_dot}), (\ref{eq:j_gw}), and (\ref{eq:jloss}), we get
%
\begin{align}
    \frac{\dot{a}}{2a}=&-\frac{32}{5}\frac{G^3}{c^5}\frac{m_1m_2M_t}{a^4}-\frac{\dot{m}_1}{m_1} \left ( \frac{m_1j_1}{J_{\rm orb}} +1\right )  -\frac{\dot{m}_2}{m_2} \nonumber \\ &+ \frac{k_1m_1R_1^2}{J_{\rm orb}\tau_{s,1}}(\omega_1-\omega_o)  +\frac{k_2m_2R_2^2}{J_{\rm orb}\tau_{s,2}}(\omega_2-\omega_o) +\left ( \frac{1}{2} -q \right )\frac{\dot{M}_t}{M_t}. \label{eq:adot}
\end{align}
We obtain the corresponding equation for the GW frequency at which the binary emits, combining $f_{\rm GW}=\frac{\omega_o}{\pi}$ with Eq.~\eqref{eq:adot}:
%
\begin{align}
    \frac{\dot{f}_{\rm GW}}{f_{\rm GW}}=&\frac{96}{5}\frac{G^3}{c^5}\frac{m_1m_2m}{a^4}+\frac{3\dot{m}_1}{m_1} \left ( \frac{m_1j_1}{J_{o}} +1\right )+\frac{3\dot{m}_2}{m_2} \nonumber \\  &-\frac{3k_1m_1R_1^2}{J_{o}\tau_{s,1}}(\omega_1-\omega_o) -\frac{3k_2m_2R_2^2}{J_{\rm orb}\tau_{s,2}}(\omega_2-\omega_o) +(3q-1)\frac{\dot{M}_t}{M_t}.\label{eq:fdot}
\end{align}

Finally, we need to specify a mass-radius relation for WDs. We take the commonly used model by Eggleton, defined in Eq. (15) of~\cite{1988ApJ...332..193V}. We note that for small mass WDs ($\lesssim 0.01 M_{\odot}$) finite-entropy effects are expected to become relevant \citep{Deloye:2003tb}. The systems we evolve do not reach this regime, so we neglect such effects. Thus defined, our model has two sets of unspecified parameters: the tidal synchronization timescale of WDs ($\tau_{s,i}$) and the spin of WDs after formation ($\omega_{i,0}$). Indeed, our population synthesis code predicts the white dwarf masses and separation at the formation of the binary, but not their spins.

As a general rule, \cite{10.1093/mnras/207.3.433} found that the tidal synchronization timescale in a binary scales as:
\begin{equation}
\tau_{s,i} \propto \left ( \frac{m_i}{m_{-i}} \right )^2 \left ( \frac{a}{R_i} \right ) ^6. \label{eq:sync_timescale}
\end{equation}
where $m_{-i}$ is the mass of the other binary component. Thus, we define a reference value $\tau_{s,{\rm ref}}$ for the heavier WD in a reference system with $m_1=0.6M_{\odot}$ and $m_2=0.25 M_{\odot}$ at the moment it overfills its Roche lobe, and use the scaling relation of Eq.~\eqref{eq:sync_timescale} to compute it at all times for any other WD in a binary. In particular, the synchronization timescale of the lighter WD in a binary is:
\begin{equation}
    \tau_{s,2}=\tau_{s,1}q^4 \left(\frac{R_1}{R_2} \right)^6.\label{eq:ts2}
\end{equation}
This approach provides a universal representation of tidal torques among DWDs, whose strength is captured by $\tau_{s,{\rm ref}}$. To explore the impact of this parameter, we consider 10 values of $\tau_{s,{\rm ref}}$ log-uniformly distributed between $10^{-2}  \ {\rm yr}$ and $10^{16} \ {\rm yr}$. The higher limit is motivated by the value of $10^{15} \ {\rm yr}$ considered in \cite{Marsh:2003rd}, \cite{2015ApJ...806...76K} and \cite{Biscoveanu:2022sul}, but, as we comment in Sec.~\ref{sec:pop}, we find results to be the same for $\tau_{s,{\rm ref}}\gtrsim 10^{8} {\rm yr}$. There remains a large uncertainty regarding the spin of WDs. Observations with the {\it Kepler} telescope have revealed a handful of slowly rotating WDs with rotation periods of hours to tens of hours~\citep{Hermes_2017,2024MNRAS.tmp..313H}. On the other hand, a few rapidly spinning WDs with rotation periods below $40 {\rm s}$ have been observed in cataclysmic variables \citep{2020MNRAS.499..149A,2020ApJ...898L..40L,2022MNRAS.509L..31P}, which suggests that WDs can reach high spins through accretion. To bracket uncertainties, we consider three prescriptions for the initial spins:
\begin{itemize}
    \item $\omega_{1,0}=\omega_{2,0}=0$, 
    \item  $\omega_{1,0}=\omega_{2,0}=\omega_o$, 
    \item $\omega_{1,0}=0.8\omega_{K,1}, \  \omega_{2,0}=0.8\omega_{K,2}$, 
\end{itemize}
where $\omega_{i,K}$ is the break-up rate defined in Eq.~\eqref{eq:breakup}. The first prescription represents slowly spinning WDs, the second assumes they are synchronised with the orbit since the formation of the binary, and the third prescription mimics rapidly spinning WDs (periods of $\sim 10-100 {\rm s}$, in agreement with the observations of rapidly rotating WDs). Although it might not be realistic to assume that all WDs rotate rapidly at birth, exploring this scenario gives us insights into observable signatures of such type of systems. Alternatively, we could consider a scenario where only the heavier WD is highly spinning, and the lighter has virtually zero spin. It turns out that the evolution of a given binary is mostly sensitive to the spin of the heavier WD, due to its much shorter tidal synchronization timescale [see Eq.~\eqref{eq:ts2}]. Thus, we expect the results in such a scenario would be qualitatively similar to the ones we obtain in the high-spin case, as we argue in Sec.~\ref{sec:syst}.

\section{Example of evolution} \label{sec:exs}

We computed the evolution of DWDs through numerical integration of Eqs.~(\ref{eq:dm}), (\ref{eq:dm1}), (\ref{eq:omega}), and (\ref{eq:adot}) with the \texttt{odeint} package of \texttt{scipy}~\citep{scipy}. For accreting systems, the evolution can be very rapid, and the integration might fail. When this happened, we resumed the integration with a higher time resolution and kept increasing the resolution until the integration was successful. For the last stage, we decreased the resolution, in order to speed up the computation.

\begin{figure}
\centering
 \includegraphics[width=0.49\textwidth]{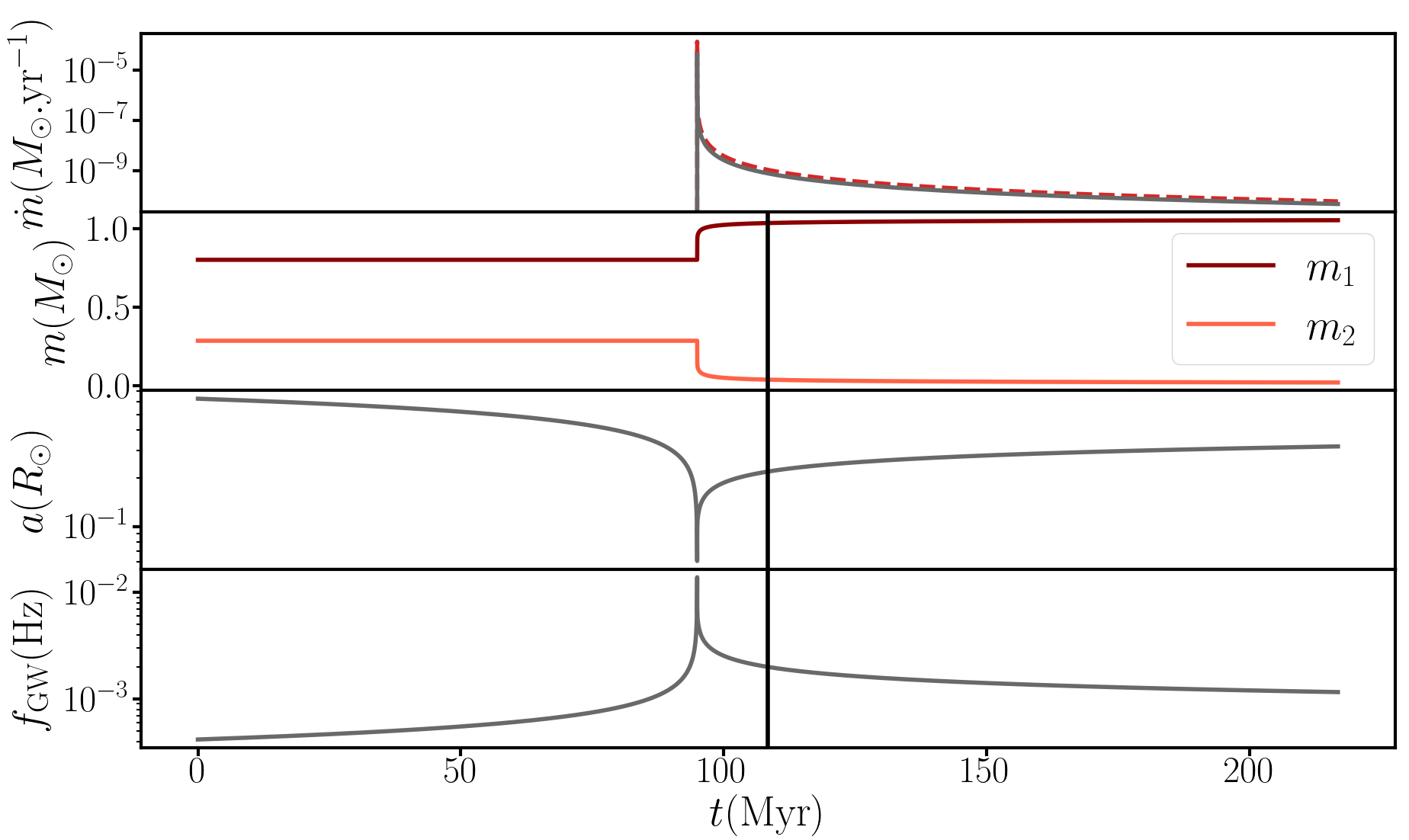}\\
 \centering
 \caption{Evolution for a system with  $m_{1,0}=0.80 M_{\odot}$ and $m_{2,0}=0.29 M_{\odot}$, and $\tau_{s,{\rm ref}}=10^{4}\ {\rm yr}$, $\omega_{1,0}=\omega_{2,0}=0$. The black vertical line shows the time the binary has to evolve between its formation and today. To show its long-term trajectory, we extend the evolutionary simulation beyond this timeframe. 
 }\label{fig:ex_ev}
\end{figure}

\begin{figure}
\centering
 \includegraphics[width=0.49\textwidth]{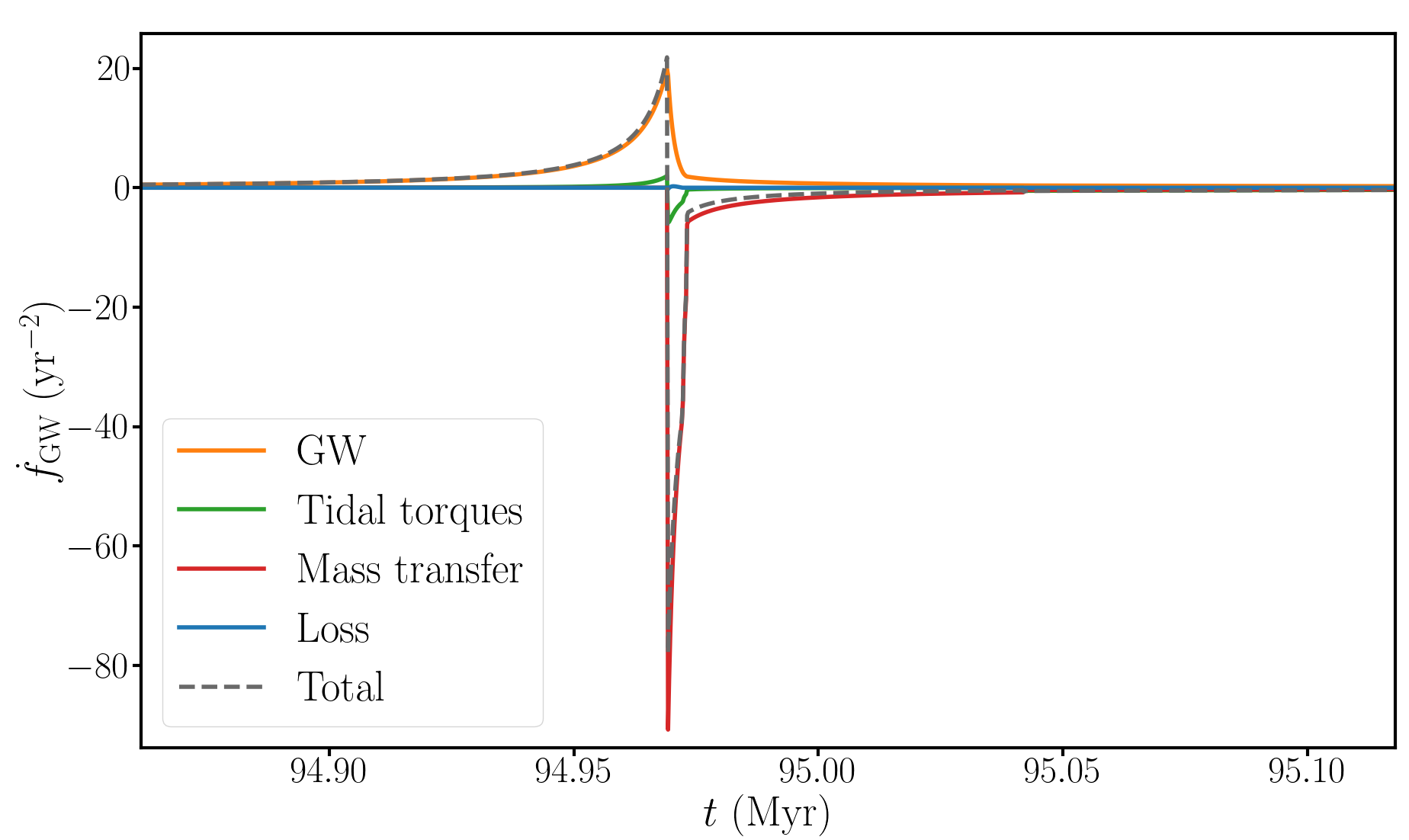}\\
 \centering
 \caption{Break-up of the contributions to $\dot{f}_{\rm GW}$ from the different terms in Eq.~\eqref{eq:fdot} for the system shown in Fig.~\ref{fig:ex_ev}, zoomed in around the moment mass transfer begins. We have combined the donor and accretor contributions into a single term for the mass transfer and for the tidal torque term. 
 }\label{fig:ex_fdot}
\end{figure}

Typically, the separation of DWDs first shrinks due to GW radiation, and, if the binary evolves for long enough, the separation becomes sufficiently small for the lighter WD to overfill its Roche lobe, and mass transfer begins. The binary then undergoes a violent accretion episode that may cause a rapid merger~\citep{2015ApJ...805L...6S,Pakmor:2022lwn}. If, instead, the binary survives this episode, it starts outspiralling. Following previous works~\citep{Webbink:1984ti,1985ApJ...297..531N,Marsh:2003rd,Kremer:2017xrg,Biscoveanu:2022sul}, we consider that a system becomes unstable, triggering a rapid merger, if at any point $-\dot{m}_d>10^{-2}M_{\odot}.{\rm yr}^{-1}$. 

We show in Figs.~\ref{fig:ex_ev} and \ref{fig:ex_fdot} the evolution of a DWD in our catalogue that undergoes stable mass transfer. For that system, $\tau_{s,{\rm ref}}=10^{4}\ {\rm yr}$ and $\omega_{1,0}=\omega_{2,0}=0$. The initial masses are $m_{a,0}=0.80 M_{\odot}$ and $m_{d,0}=0.29 M_{\odot}$ and the final masses $m_{1,f}=1.03 M_{\odot}$ and $m_{2,f}=0.04M_{\odot}$. The vertical black line shows the time from the formation of the binary until now. In the plots shown in this section, we have evolved the systems for longer to show the long-term evolution of accreting DWDs. Fig.~\ref{fig:ex_ev} shows the main features of the evolution of accreting systems: after the lighter WD overfills its Roche lobe, the accretion rate increases rapidly and, almost immediately, causes the distance between WDs to increase (negative $\dot{f}_{\rm GW}$), reducing the orbital frequency back to a few mHz and leading to a drop-off in the accretion rate. The binary then evolves slowly in an equilibrium configuration. This equilibrium is characterised by $\dot{\Delta}=0$. Deriving Eq.~\eqref{eq:overfill}, we have
\begin{equation}
    \dot{\Delta}=\left ( \zeta_2 R_2 -\zeta_{r_L}R_L \right ) \frac{\dot{m}_2}{m_2}-R_L\frac{\dot{a}}{a},
\end{equation}
with $\zeta_2=\frac{{\rm d} \log(R_2)}{{\rm d}\log(m_2)}$ and $\zeta_{r_L}=\frac{{\rm d} \log(R_L/a)}{{\rm d}\log(m_2)}$. Setting $\dot{\Delta}=0$ in the above equation and using Eq.~\eqref{eq:adot}, we get the mass transfer rate at equilibrium:
\begin{equation}
    \dot{m}_{2,{\rm eq}}=\frac{\frac{32}{5}\frac{G^3}{c^5}\frac{m_1m_2m}{a^4}}{1-q-\frac{m_2j_1}{J_{\rm orb}}+\frac{1}{2}
    \left( \zeta_2 (R_2/R_L) -\zeta_{r_L} \right ) }.
\end{equation}
From the upper panel of Fig.~\ref{fig:ex_ev}, we can see that the equilibrium mass loss rate matches very well the actual rate. The small discrepancy is due to our neglect of sub-dominant tidal terms. We have verified that when we remove those terms from the evolution equation [Eq.~\eqref{eq:adot}], the match is perfect. From the second and fourth panels of Fig.~\ref{fig:ex_ev}, we can see that most of the change in mass and frequency occurs at the very beginning of the mass transfer phase, within a few thousand years, after which they change little. The binary system loses mass only during this short time window. Subsequently, $\dot{m}_2$ drops by several orders of magnitude, falling below the Eddington rate, such that $\dot{M}_t=0$. The work of \cite{Yi:2023osk} exploits the existence of such a simple expression for the mass-transfer rate at equilibrium, which can be written in terms of the instantaneous properties of the binary, to obtain measurements of the individual masses for accreting DWDs. In Appendix.~\ref{app:res}, we discuss a case occurring in specific conditions, where the donor successively overfills and underfills its Roche lobe, leading to very pronounced oscillations in the evolution. 


 \begin{figure*}[h!]
\centering
 \includegraphics[width=0.88\textwidth]{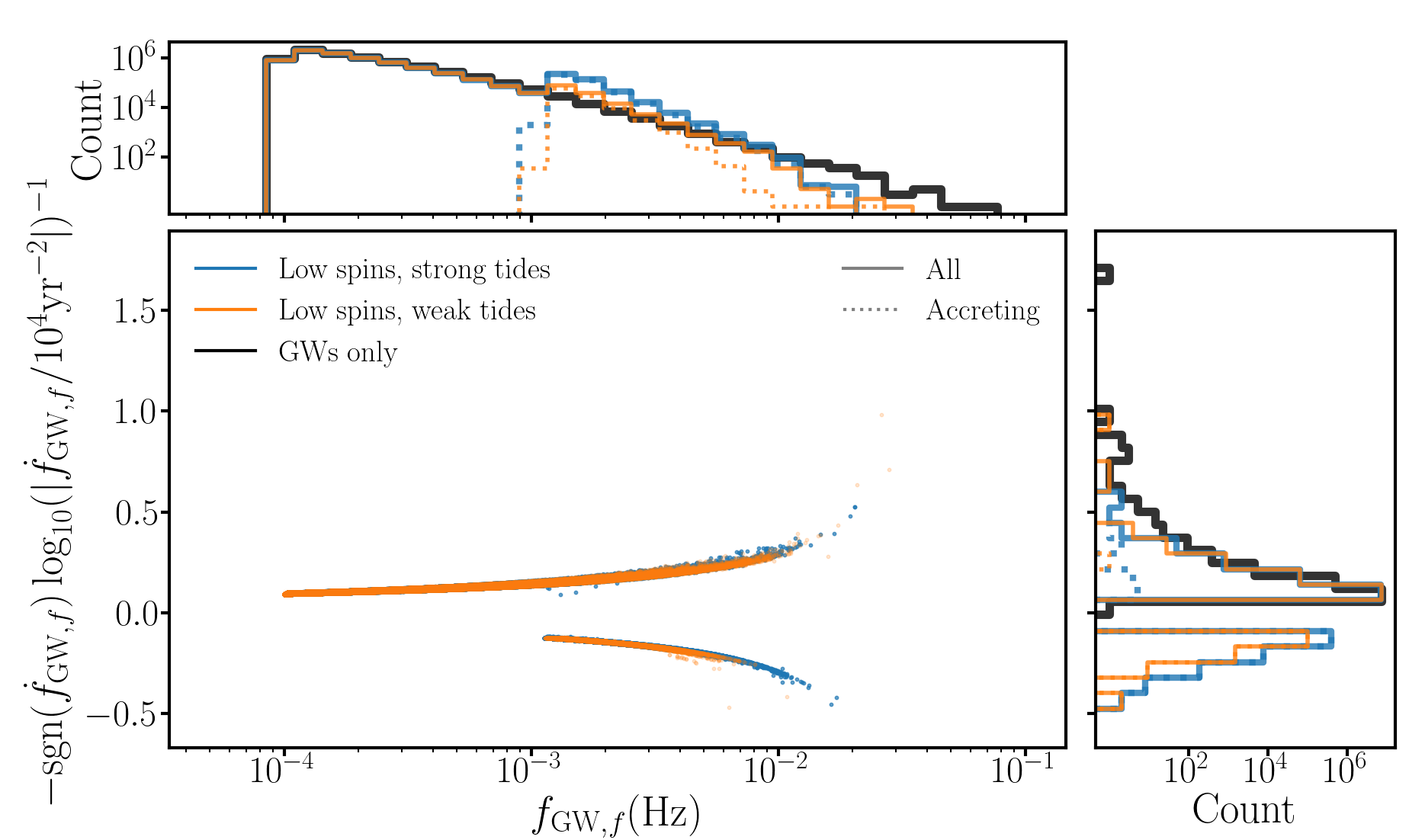}\\
 \centering
 \caption{Comparison of the populations in the $(f_{{\rm GW},f},F(\dot{f}_{{\rm GW},f}))$ plane for the 'low spins, strong tides' (in blue) and 'low spins, weak tides' (in orange) scenarios, where $ F(\dot{f}_{{\rm GW},f}))=-{\rm sgn}(\dot{f}_{{\rm GW},f})\log_{10}(|\dot{f}_{{\rm GW},f}/10^4 {\rm yr}^{-2}|)^{-1}$. This quantity has the same sign and monotonic behaviour as $\dot{f}_{{\rm GW},f}$. The black lines in the upper and right panels show the result in the 'GWs only' scenario. Accreting systems are typically those with $\dot{f}_{{\rm GW},f}<0$. We recall that we keep only DWDs that have $f_{{\rm GW},f}\geq 0.1 \ {\rm mHz}$ in the 'GWs only' case, which explains why the distribution goes to 0 below this limit and for small positive values of $\dot{f}_{{\rm GW},f}$.}\label{fig:comp_dfgwf_fgwf1}
\end{figure*}

 \begin{figure*}[h!]
\centering
 \includegraphics[width=0.88\textwidth]{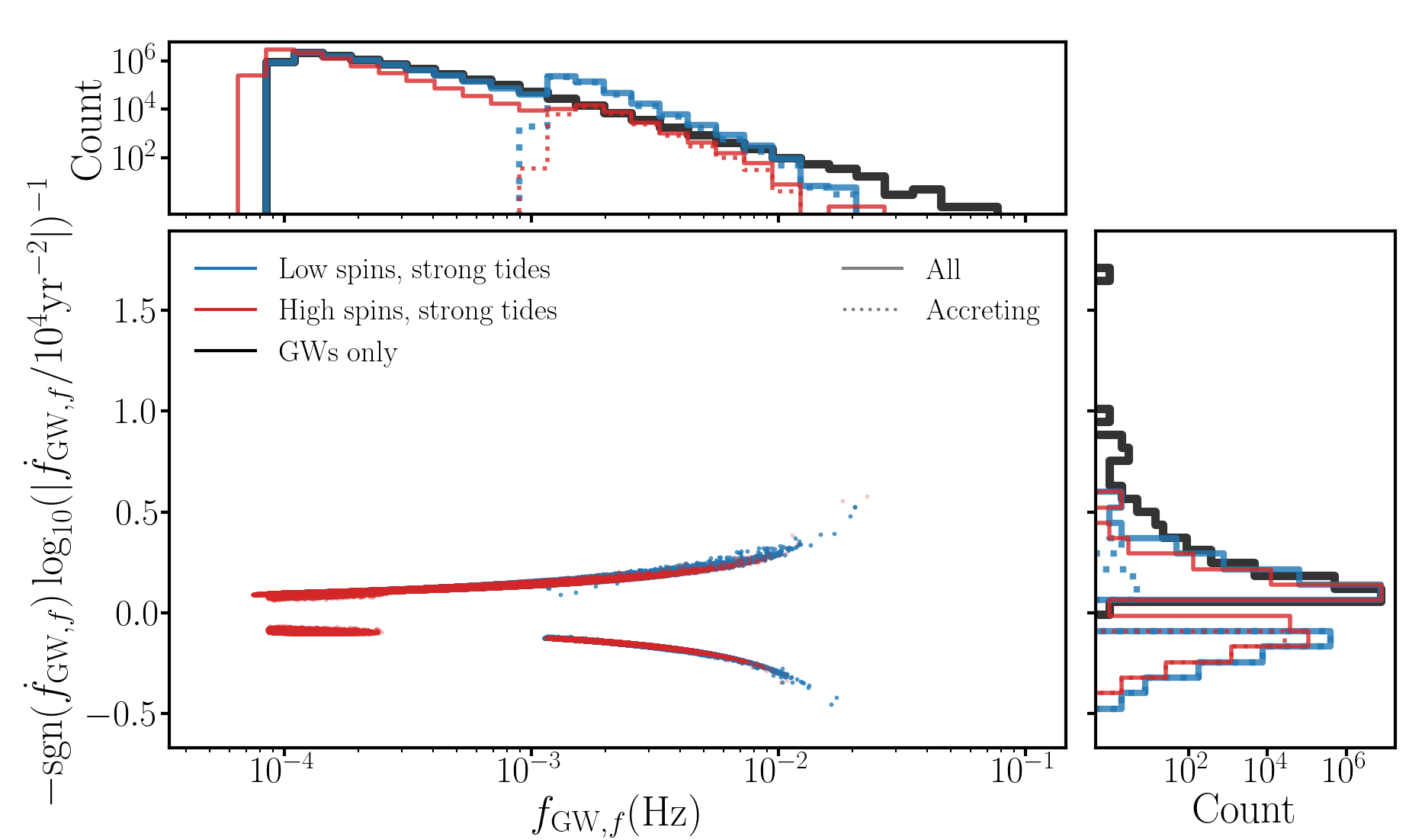}\\
 \centering
 \caption{Same as Fig.~\ref{fig:comp_dfgwf_fgwf1}, but for the 'low spins, strong tides' (in blue) and 'high spins, strong tides' (in red) scenarios. For WDs formed with high spins, tidal torques can cause the binary to outspiral even before the onset of mass transfer, explaining why some systems in red have $f_{{\rm GW},f}< 0.1 \ {\rm mHz}$.}\label{fig:comp_dfgwf_fgwf2}
\end{figure*}

As shown in Fig.~\ref{fig:ex_fdot}, the initial evolution is driven by GW radiation and later by mass transfer. Tidal torques initially accelerate the shrinking of the orbit. Once the WDs start exchanging mass, the heavier WD is spun up, and $\omega_{1}>\omega_o$. The tidal torque on the heavier WD changes sign and helps to pull the binary apart. Thus, strong enough tidal torques can prevent binaries from getting too close and prevent them from merging rapidly.

The evolution of binaries with initially wider separations, such that they do not reach small enough separations for the donor to overflow its Roche lobe, is very similar to the early stages of the system shown in Fig.~\ref{fig:ex_ev}. The separation diminishes slowly, accompanied by a slow increase in the GW frequency, while the masses of the components remain unchanged. A noteworthy exception to this scenario occurs when WDs are born with large spins and tidal synchronization is very efficient: the binary can outspiral while far apart, without mass transfer. This happens because, for highly spinning WDs, tidal torques give a positive contribution to $\dot{a}$ [see Eq.~\eqref{eq:adot}] that can be large enough to overcome GW radiation. We discuss the implications of this in the next section.

\section{Results for the population of DWDs}\label{sec:pop}

In this section, we present the results for the whole DWD population. We excluded systems where: the accretor exceeds the Chandrasekhar mass, or $-\dot{m}_d>10^{-2}M_{\odot}.{\rm yr}^{-1}$, or the separation goes below $R_1+R_2$, as the latter is interpreted as a merger. We focus on three combinations of tidal synchronization timescales and initial spins among the thirty considered in this work, and give them the following labels:
\begin{enumerate}
    \item 'low spins, strong tides': $\omega_{1,0}=\omega_{2,0}=0, \ \tau_{s,{\rm ref}}=1{\rm yr}$,
    \item 'low spins, weak tides': $\omega_{1,0}=\omega_{2,0}=0, \ \tau_{s,{\rm ref}}=10^{12}{\rm yr}$, 
    \item 'high spins, strong tides': $\omega_{1,0}=0.8 \omega_{K,1}, \ \omega_{2,0}=0,8\omega_{K,2}, \ \tau_{s,{\rm ref}}=1{\rm yr}$. 
\end{enumerate}
Moreover, we denote by 'GWs only' the case where DWDs evolve solely through GW radiation. For the sake of limiting computational power, we restrict ourselves to binaries with $f_{{\rm GW},f}$ above $0.1 {\rm mHz}$ in the 'GWs only' case.

In Fig.~\ref{fig:comp_dfgwf_fgwf1}, we show the present-day derivative of the GW frequency as a function of the final frequency itself in the 'low spins, strong tides' (in blue) and 'low spins, weak tides' (in orange) scenarios. In the upper and right panel we also plot the histograms for the 'GWs only' case. To make the figure more readable, we plot $F(\dot{f}_{{\rm GW},f})=-{\rm sgn}(\dot{f}_{{\rm GW},f})\log_{10}(|\dot{f}_{{\rm GW},f}/10^4 {\rm yr}^{-2}|)^{-1}$ rather than $\dot{f}_{{\rm GW},f}$. The minus sign and the $10^4 {\rm yr}$ normalisation are so that systems with positive (negative) $\dot{f}_{{\rm GW},f}$ are shown as positive (negative) for the other quantity, and the inversion is so that it has the same monotonic behaviour as $\dot{f}_{{\rm GW},f}$. We show with dotted lines the contribution of accreting systems to the total histograms. Those systems typically have $\dot{f}_{{\rm GW},f}<0$. A handful of those, at most a few tens, might have positive $\dot{f}_{{\rm GW},f}$ because they just started the accretion process, and did not have time to start outspiralling, or because they are undergoing oscillations (see Appendix.~\ref{app:res}).

In the $f_{{\rm GW},f}$ histogram (upper panel), we observe a bump due to the accumulation of outspiralling accreting binaries around a few mHz, with a peak at $\sim 1.3 \ {\rm mHz}$. This is not a natural limit of the evolution, but rather a typical value around which DWDs accumulate given the time they have to evolve. As described in Sec.~\ref{sec:exs}, the late evolution of systems that survive the violent accretion episode at the onset of mass transfer is very slow, leading to the stacking of systems at a few mHz. Strong tidal torques help stabilise the evolution and prevent rapid mergers. Therefore, we observe that, although the bump is also observable for large values of $\tau_{s,{\rm ref}}$, it is more prominent for shorter tidal synchronization timescales: we find 10 times more systems at 1.3 mHz relative to the 'GWs only' scenario in the 'low spins, strong tides' case and 3 times more in the 'low spins, weak tides' case. The lack of events at high frequency compared to the 'GWs only' scenario is because binaries that reach high frequencies either merge rapidly or outspiral. Therefore, this feature is very sensitive to our criterion for the stability of mass transfer, and it is possible that this is a failure of our model. Finally, we note that the low frequency portion of the population (below 1mHz) is pushed to higher frequencies due to tidal torques that accelerate the evolution of the systems (positive contribution to $\dot{f}_{\rm GW}$, see Fig.~\ref{fig:ex_fdot}), with a more pronounced effect in the 'low spins, strong tides' case. 

Next, we compare the 'low spins, strong tides' (in blue) and 'high spins, strong tides' (in red) scenarios on Fig.~\ref{fig:comp_dfgwf_fgwf2}. In the latter scenario, the initial spins of the WDs are much larger than the orbital frequency, and tidal torques give a large positive contribution to $\dot{a}$ even without mass transfer [see Eq.~\eqref{eq:adot}]. Therefore, tidal torques tend to slow down the evolution, and in some cases they can even dominate over GW radiation, leading the binary to outspiral. This is why, in that case, some systems are below 0.1 mHz, although the original population we evolve contains only binaries that have $f_{{\rm GW},f}>0.1 \ {\rm mHz}$ in the 'GWs only' case. Some of the binaries below 0.1 mHz have positive frequency derivative because GW radiation ended up taking over again, causing them to inspiral. If a handful of DWDs were detected at low-frequency ($\lesssim$ 0.1 mHz) with negative $\dot{f}_{{\rm GW},f}$, it would suggest that those are highly-spinning WDs and that tidal synchronization is efficient within the binary. However, as we show in Sec.~\ref{sec:da}, it might not be trivial to detect confidently such a subpopulation given the small values of $|\dot{f}_{{\rm GW},f}|$ they have and the low signal-to-noise ratio (S/N) of sources at low frequency. We stress that for this subpopulation to form, it is not necessary that all WDs have large spin. Finally, the peak around a few mHz is less prominent in the high-spins scenario than in the low-spins one because of the slow-down of the evolution due to tidal torques: some systems do not have time to start accreting, whereas they do in the low-spins scenario. 

\begin{figure}
\centering
 \includegraphics[width=0.49\textwidth]{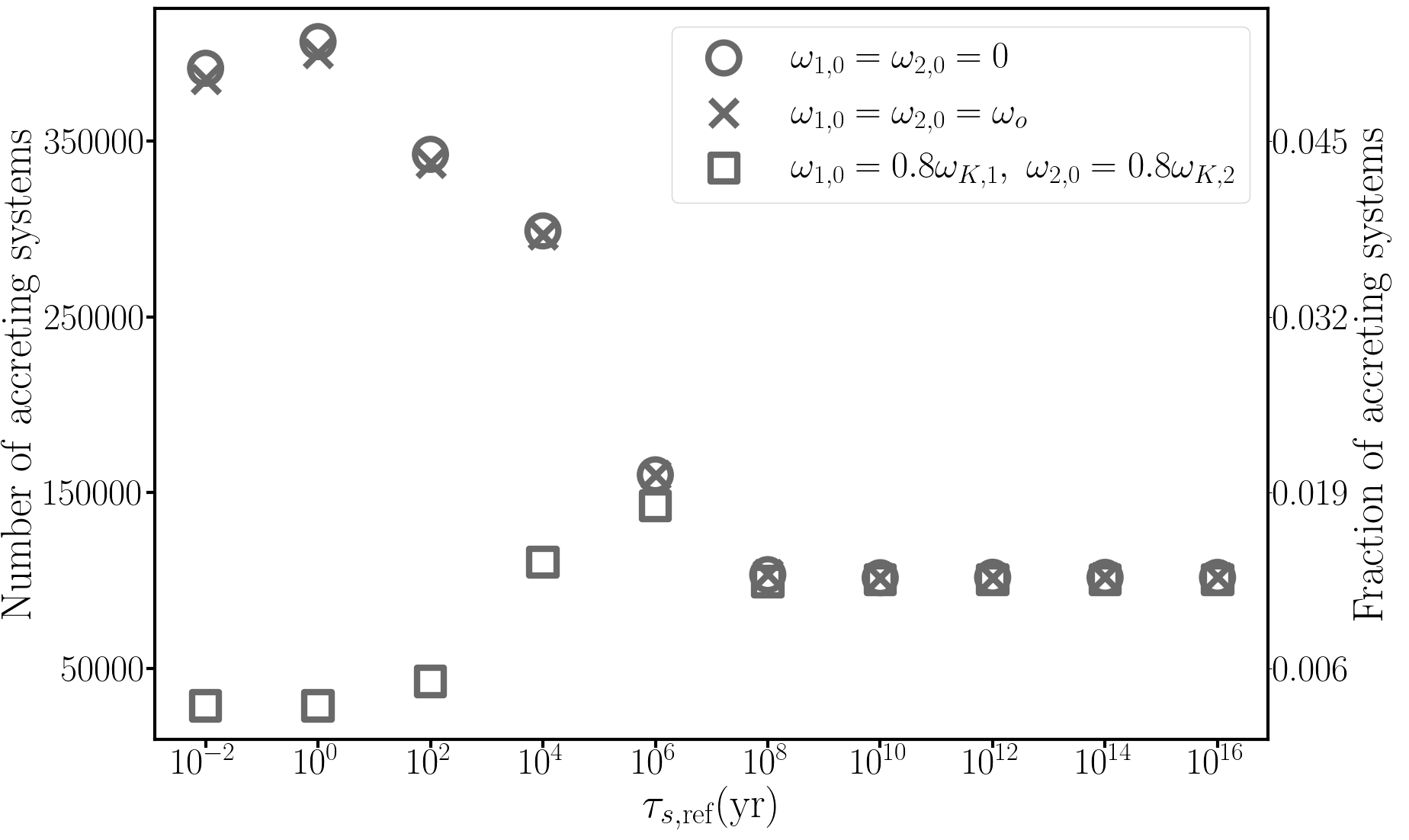}\\
 \centering
 \caption{Number of accreting systems across all configurations under consideration. The left y-axis denotes the absolute count, while the right y-axis indicates the corresponding fraction relative to the initial catalogue. The DWD catalogue used as a starting point contains 7740598 systems.
 }\label{fig:comp_naccretion}
\end{figure}

As $\tau_{s,{\rm ref}}$ increases, the subpopulation of low-frequency binaries with negative $\dot{f}_{{\rm GW},f}$ disappears, and the initial spin has no impact. This can be seen in Fig.~\ref{fig:comp_naccretion}, where we show the number of accreting systems for each configuration, which is closely related to the prominence of the peak around a few mHz. For $\tau_{s,{\rm ref}}\gtrsim 10^{8} {\rm yr}$, the initial spin frequency has little importance because tidal torques become negligible. For smaller tidal synchronization timescales, the number of accreting systems increases (decreases) for slowly spinning (highly spinning) WDs. When $\omega_{1,0}=\omega_{2,0}=\omega_o$, tidal torques are initially zero, but as the separation decreases, the orbital frequency increases and the WD spins have to 'catch up' with it. Therefore, they tend to be smaller than $\omega_o$, and, before the onset of mass transfer, tidal torques give a negative contribution to $\dot{a}$ [see Eq.~\eqref{eq:adot}], as in the case $\omega_{1,0}=\omega_{2,0}=0$. However, the absolute value of the contribution of tidal torques is less significant (because $\omega_o-\omega_i$ is smaller), and some systems do not have time to start accreting within the evolution time, explaining the slight decrease in the number of accreting systems. This is a small effect, and our results support that the zero spin scenario is representative of the case where WDs are formed with small spins. For weak tidal coupling, $\tau_{s,{\rm ref}}\gtrsim 10^{8} {\rm yr}$, we find $\sim 10^6$ systems to be accreting, which represents a little more than $1 \%$ of all DWDs in the catalogue. As the tidal coupling becomes stronger, the number of accreting systems increases to $\sim 4\times 10^6$, i.e. $\sim 5\%$ of all systems, for low spins, and goes down to $\sim 3 \times 10^5$, i.e. $0.4\%$ of all systems, for high spins.

\begin{figure*}
\centering
\includegraphics[width=0.45\textwidth,valign=t]{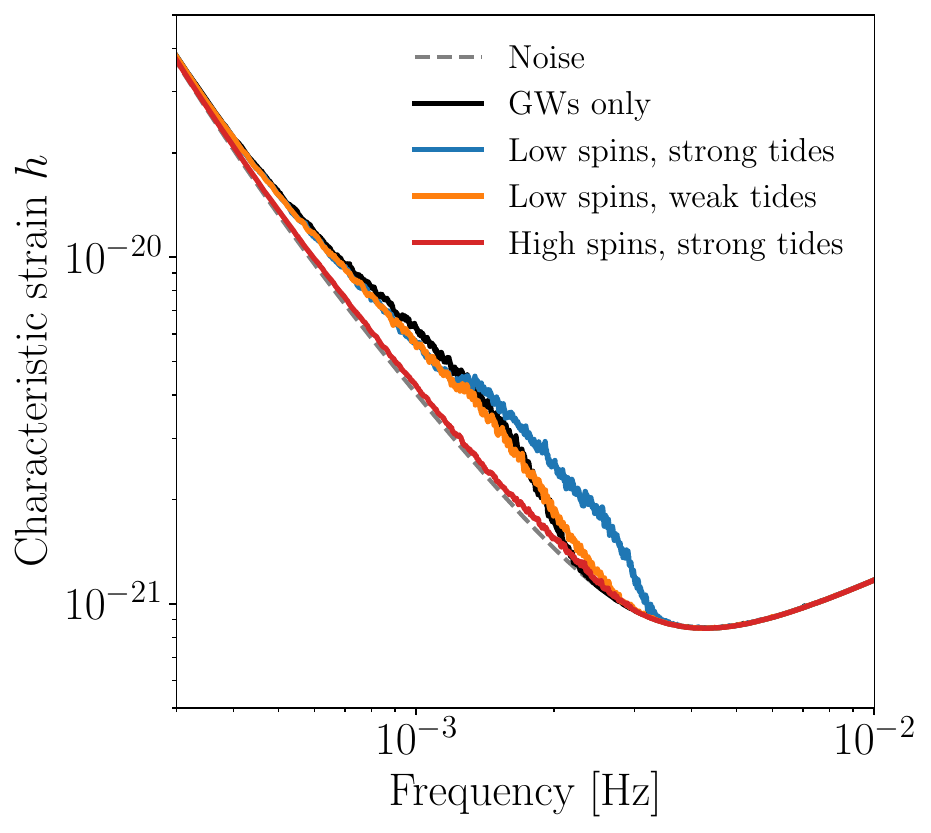}
\centering
\includegraphics[width=.3\textwidth,valign=t]{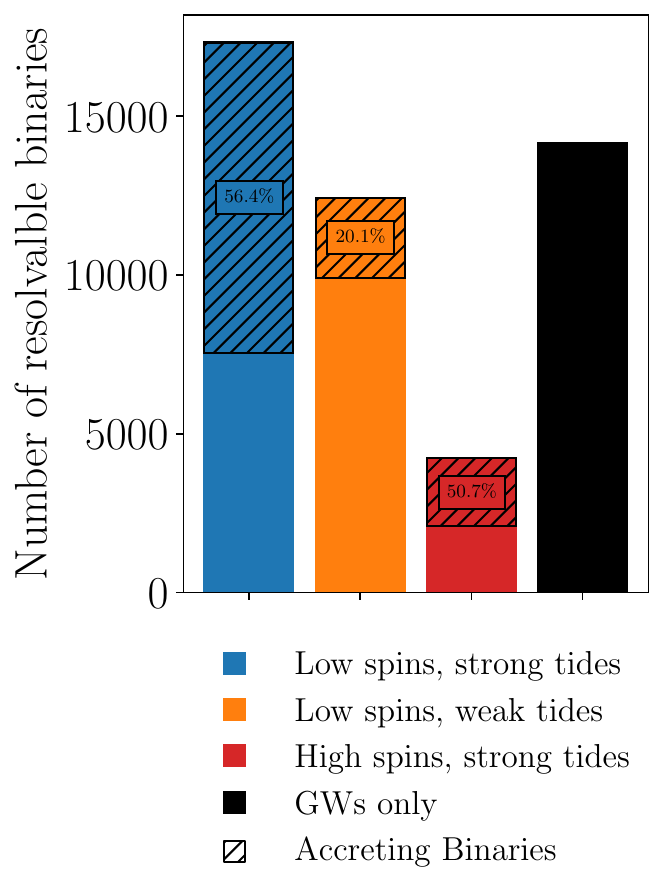}
\caption{LISA results obtained with the subtraction algorithm in different scenarios, indicated by colours. {\it Left}: Characteristic strain of the total noise power spectral density $\left (\sqrt{f S_n(f)} \right )$, including the instrumental and the DWD confusion noise, compared to the instrumental noise alone, in grey dashed lines. 
{\it Right}: Number of resolvable sources in each of the considered scenarios. Hatches indicate how many of those are accreting. We find $56.4\%$, $20.1\%$, and $50.7\%$; of the resolvable systems to be accreting in the 'low spins, strong tides', 'low spins, weak tides', and 'high spins, strong tides' scenarios, respectively, as indicated by the boxed numbers. 
}\label{fig:data_analysis}
\end{figure*}

\section{Data analysis and predictions for LISA }\label{sec:da}

Next, estimated the implications of the different scenarios discussed above for LISA observations. The GW signal emitted by DWDs is almost monochromatic, and takes the simple form

\begin{align}
h_{+}&=A_0 \frac{1}{2} \left (1+\cos^2(\iota) \right ) \cos(\phi_0+2\pi f_{{\rm GW},f}t+\pi \dot{f}_{{\rm GW},f} t^2), \\
h_{\times}&=A_0 \cos(\iota) \sin(\phi_0+2\pi f_{{\rm GW},f}t+\pi \dot{f}_{{\rm GW},f} t^2), \label{eq:gw_signal}
\end{align}
where $A_0=\frac{2\mathcal{M}_cG^{5/3}}{c^3D_L}(\pi \mathcal{M}_cf_{{\rm GW},f})^{2/3}$ gives the amplitude of the signal, $\mathcal{M}_c=m_1^3 m_2^3/M_t$ is the chirp mass of the binary, $D_L$ the luminosity distance, $\iota$ the inclination of the binary with respect to the line of sight and $\phi_0$ the initial phase. We computed LISA's response to such GW signals following~\cite{Cornish:2007if} to generate mock data, and the {\it LISA data challenge} software~\citep{ldc}. The sky location, which enters the LISA response, and the distance to the source were provided in the catalogue. We drew the inclination angle and the initial phase assuming isotropic orientation. We took a flat distribution for the polarisation angle, which is a quantity that enters the LISA response. We assumed a $4 \ {\rm yr}$ LISA mission and took the SciRDv1 curve~\citep{scirdv1} for the instrumental noise.

The detection and characterisation of individual sources in the LISA data is difficult~\citep{Littenberg:2023xpl,Strub:2023zxl,Zhang:2021htc}. The challenge can be attributed mostly to the very high number of different types of sources, which is expected to cause a high level of overlap between their measured GW signatures in the LISA band. For that reason, there is significant effort focusing on the developing Global Fit data analysis pipelines \citep{Littenberg:2023xpl}, which are based in algorithms that can simultaneously infer the parameters of each binary, as well as the actual number of binary signatures present in the data streams. As expected, these analyses are complex and often time and energy consuming. Thus, in order to characterise the number of resolvable binaries and the remaining confusion noise, we use a different approach.

Our analysis is based on the methodology presented in~\cite{Karnesis:2021tsh}, \cite{tim06}, \cite{cro07}, \cite{nis12} and \cite{Lamberts:2019nyk}, which is based on an iterative procedure that classifies the sources as resolvable based on S/N criteria. The process begins ($k=0$) with an initial estimate of the overall noise power spectral density $S_{n,k}$, which is the sum of the instrumental component and the confusion noise signal generated by the given population of signals. The initial estimate of the noise can be performed using different recipes, for example via a running median on the power spectral density of the data. Then, we iterate over the catalogue of sources, and classify the ones with S/N above a given threshold as resolvable. For our analysis, we chose this threshold to be equal to ${\rm S/N}_\mathrm{thres} = 7$. The resolvable sources are subtracted from the data, and a new iteration begins by re-estimating the overall noise $S_{n,k+1}$. The algorithm stops when no more sources can be subtracted, which can either mean that all sources in the catalogue are resolved, or that all the remaining sources are below the given threshold.

    This algorithm is useful because it yields results quickly, without requiring a complete Bayesian trans-dimensional analysis. At the same time, it employs different simplifying assumptions, such as perfect data sets (no noise transients of missing data), minimal noise correlations, and perfect residuals (sources are subtracted using their true properties, which ignores uncertainties and possible correlations between the sources). Considering the above, this procedure can be seen as an optimistic upper limit to the capabilities of a full Bayesian analysis on the LISA data~\citep{Karnesis:2021tsh}.

The outputs of this algorithm are summarised in Fig.~\ref{fig:data_analysis}. On the left panel, we plot the characteristic strain of the total noise power spectral density $\left (\sqrt{f S_n(f)} \right)$ , including both instrumental noise and the DWD confusion noise, in different scenarios, and on the right panel we show the number of resolvable binaries in each scenario. The hatched part of the histograms indicates the contribution of accreting binaries to all resolvable binaries. Overall, pronounced effects such as strong tides or high spins manifest observable signatures. In the 'low spins, strong tides' case, the confusion noise is significantly different from the 'GWs only' case, and in particular is much higher between 1 and 3 mHz due to accreting systems. At $ 1.3 \ {\rm mHz}$ (the peak of accreting events),$\sim 10^{-2}\%$ of accreting systems are resolvable, whereas $\sim 2\%$ of non-accreting ones at the same frequency are resolvable. This occurs because mass transfer decreases the mass ratio, and thus also the chirp mass, which reduces the GW amplitude [see Eq.~\eqref{eq:gw_signal}]. As the frequency increases, this fraction increases, and above 3 mHz, accreting systems are as likely to be resolved as non-accreting ones: almost all DWDs above that frequency are resolvable. Thus, accreting systems affect mostly the confusion noise between 1 and 3 mHz. In total, $\sim 3\%$ of accreting binaries are resolvable, whereas $\sim 20\%$ of non-accreting binaries with $f_{{\rm GW},f}>1.3 \ {\rm mHz}$ are resolvable. In the 'low spins, strong tides' scenario, we also predict a larger number of resolvable systems, because tidal torques accelerate the evolution of binaries, increasing the frequency at which LISA observes non-accreting systems, and therefore their GW amplitude [see Eq.~\eqref{eq:gw_signal}]. 

The 'low spins, weak tides' case yields results very similar to the 'GWs only' case. The bump of systems around a few mHz visible in Fig.~\ref{fig:comp_dfgwf_fgwf1} contributes to a slight increase of the confusion noise in that frequency band, and the number of resolvable systems is slightly smaller than in 'GWs only' scenario. The reason is that tidal effects are less effective at increasing the GW frequency and accreting systems mostly contribute to the confusion noise, but much less than in the 'low spins, weak tides' case, due to the smaller number of systems. The 'high spins, strong tides', in turn, predicts a much weaker confusion noise: the total noise differs only slightly from the instrumental noise. Moreover, it yields a much lower number of resolvable binaries: fewer sources are above 1~mHz, because of the slow-down due to tidal torques, and so they have lower S/N than in the previous scenarios. We stress that the assumption that WDs are rapidly spinning at birth is most likely not astrophysically realistic at the level of the whole population, so this result should be taken with caution. More relevant in that scenario, none of the low-frequency systems with negative $\dot{f}_{{\rm GW},f}$ are resolvable, leaving for future detectors the possibility of identifying this putative subpopulation. 

\begin{figure*}
\centering
\includegraphics[width=0.95\textwidth,valign=t]{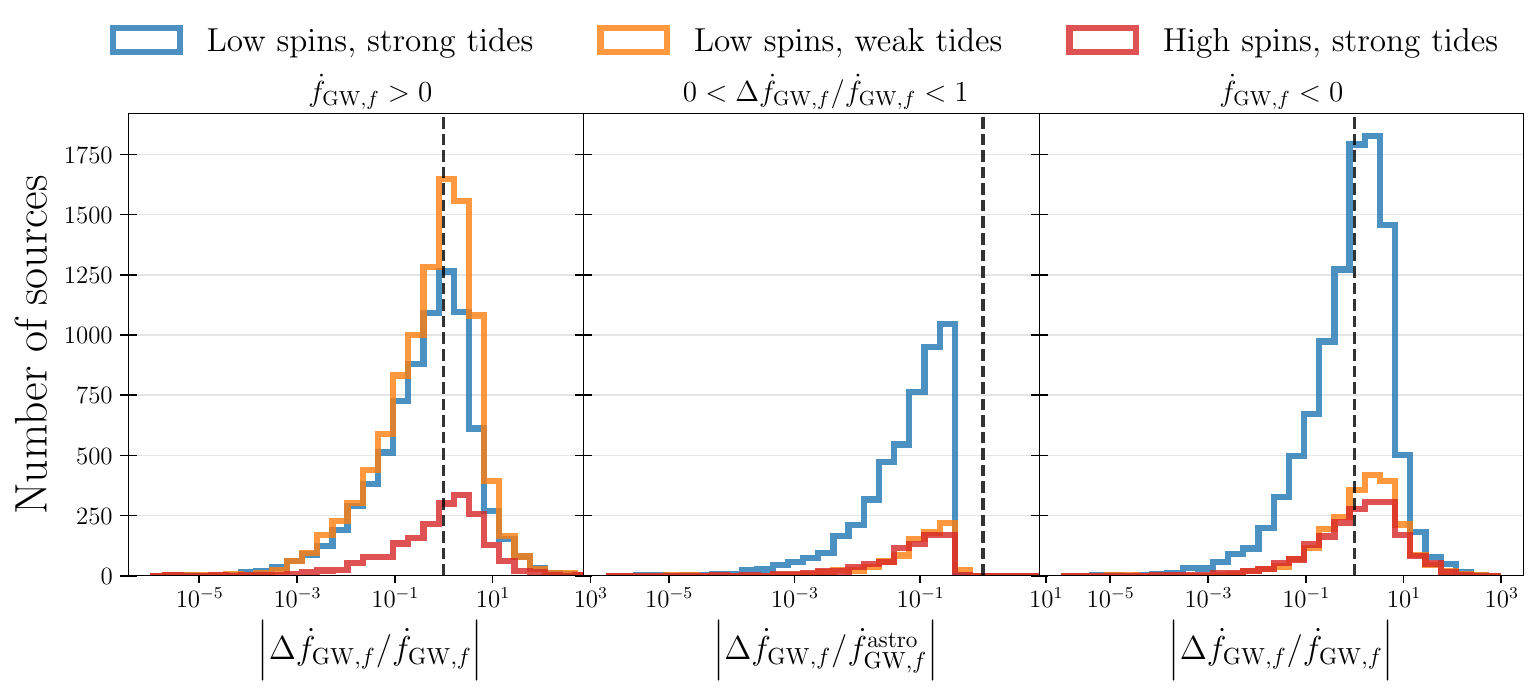}
\caption{Results of the Fisher matrix analysis. {\it Left}: Relative error on $\dot{f}_{{\rm GW},f}$ computed with the Fisher matrix for resolvable systems with $\dot{f}_{{\rm GW},f}>0$. {\it Middle}: Ratio of the measurement error to the astrophysical contribution to the total GW-frequency derivative, $f_{{\rm GW},f}^{\rm astro}$, restricted to systems for which $\dot{f}_{{\rm GW},f}>0$ and $\Delta \dot{f}_{{\rm GW},f}/\dot{f}_{{\rm GW},f}<1$. {\it Right}: Same as the left panel, but for resolvable systems with $\dot{f}_{{\rm GW},f}<0$. We consider quantities with fractional uncertainties less than 1 (i.e. to the left of the vertical black dashed line) to be measurable. } \label{fig:err_fdot}
\end{figure*}

Several science cases with DWDs rely on measuring $f_{{\rm GW},f}$ and $\dot{f}_{{\rm GW},f}$ for individual systems. 
In order to assess how many events could be used for such studies, we use the Fisher matrix~\citep{Takahashi:2002ky,Vallisneri:2007ev} to estimate the measurement error $\Delta \dot{f}_{{\rm GW},f}$ (defined here as the standard deviation predicted by the Fisher matrix) on the GW-frequency derivative for resolvable systems. In Fig.~\ref{fig:err_fdot}, we show the fractional error on $\dot{f}_{{\rm GW},f}$, splitting between accreting (left panel) and non-accreting (right panel) systems. We adopt the simple criterion $\lvert \Delta \dot{f}_{{\rm GW},f}/\dot{f}_{{\rm GW},f}\rvert <1$ to decide on the measurability of $\dot{f}_{{\rm GW},f}$ for a given system. 
In the left panel, we observe that approximately half of the resolvable systems that are not accreting have measurable GW-frequency derivative, $\sim 5000$ systems for the cases with low spins and $\sim 1000$ for the cases with high spin. However, proposed ways to measure the chirp mass and to perform tests of GR for such events rely on the evolution of DWDs being mainly driven by GWs. Even for non-accreting systems, $\dot{f}_{{\rm GW},f}$ might receive significant non-GW contributions, as we now argue. Following the notation introduced in~\cite{Breivik:2017jip} and \cite{Kremer:2017xrg}, we split the GW-frequency derivative into a purely gravitational (GR) and an astrophysical contribution as
\begin{align}
\dot{f}_{\rm GW}^{{\rm GR}}&=\frac{96}{5}\frac{f_{\rm GW}^2}{\pi c^5} \left (\pi G \mathcal{M}_c f_{\rm GW} \right )^{5/3},
\label{eq:fdto_gw}\\
\dot{f}_{\rm GW}^{{\rm astro}}&=\dot{f}_{\rm GW}-\dot{f}_{\rm GW}^{{\rm GR}}. 
\end{align}
We might interpret $\dot{f}_{\rm GW}^{{\rm astro}}$ as a systematic error, and consider it as relevant if it is larger than the statistical error. In the middle panel of Fig.~\ref{fig:err_fdot}, we show the ratio between the statistical error and $\dot{f}_{{\rm GW},f}^{{\rm astro}}$, restricting ourselves to systems with measurable positive GW-frequency derivative. In all three scenarios, for most of the considered systems $\lvert \Delta \dot{f}_{{\rm GW},f}/\dot{f}_{{\rm GW},f}^{\rm astro} \rvert <1$, indicating that neglecting astrophysical effects would significantly bias the estimation of chirp mass. This occurs because systems with measurable GW-frequency derivative are in the high-frequency end of the population, where tidal torques are stronger [see Fig.~\ref{fig:ex_fdot}]. Within our model, in the 'low spins, strong tides' and 'low spins, weak tides' cases, we necessarily have $\dot{f}_{\rm GW}^{{\rm astro}}>0$ for the systems considered here, and the chirp mass measurement would be biased to larger values [see Eq.~\eqref{eq:fdto_gw}], see also~\cite{Fiacco:2024ywf}. 
In the 'high spins, strong tides' case, we would instead underestimate the chirp mass. Regarding accreting systems, we find that the GW-frequency derivative is measurable for approximately half of the systems, $\sim 5000$ for the cases with strong tides and $\sim 1000$ for the cases with weak tides. Such systems could be used for the investigations suggested in~\cite{Breivik:2017jip} and \cite{Yi:2023osk}. We stress that those approaches rely, at least to some extent, on the modelling of astrophysical effects in DWDs, which therefore introduces a possible source of systematic error. On the other hand, similar approaches, for example establishing universal relations, could be used to exploit the measurement of resolvable non-accreting systems. Finally, the Fisher matrix approximation is not expected to hold exactly for all the systems in the catalogue, possibly yielding significant differences with full Bayesian analyses, but we expect our conclusions at the level of the population to hold.



\section{Model systematics}\label{sec:syst}

Given the amount of uncertainty regarding the evolution of DWDs, we test the robustness of our conclusions by considering different hypotheses in the modelling of DWDs evolution. We do not explore here the impact of modifications to the initial DWD population. Uncertainties in the star formation rate, stellar evolution and binary interactions can strongly impact properties of the population in terms of global numbers, orbital parameters, masses and positions. This will affect their later evolution and detectability by LISA \citep{LISA_astroWP}.

First, we consider the approach of~\cite{Gokhale:2006yn}. It is very similar to our fiducial model presented in Sec.~\ref{sec:formalism}, except that the specific angular momentum of the donor is not 0, but $j_2=R_2\omega_2^2$, and the mass lost by the system is assumed to have the same specific angular momentum as the accretor, so that $\dot{J}_{\rm loss}=\dot{M}_t j_1$. We find very similar results to our fiducial case. Slightly fewer systems survive accretion, due to less angular momentum being lost by the binary (smaller $\dot{J}_{\rm loss}$), but the main difference is that the independency on the spins for large tidal synchronization timescales, observed on Fig.~\ref{fig:comp_naccretion}, no longer holds, though the difference between different initial spin scenarios is still small. This happens because the loss of angular momentum by the donor helps stabilise the evolution. Mathematically, the term $\frac{\dot{m}_2}{m_2}$ in Eq.~\eqref{eq:adot} is now replaced by $\frac{\dot{m}_2}{m_2} \left ( \frac{m_2j_2}{J_{\rm orb}} +1\right )$, and thus it gives a more important positive contribution to $\dot{a}$, pushing the binary away, and preventing it from becoming unstable. Overall, our results confirm that the spin of the lighter WD and the model used for its specific angular momentum play a small role, unless $j_2$ grows with the spin and the lighter WD is rapidly spinning. Astrophysically, it is expected that the heavier WD could be rapidly spinning, due to the accretion of from the companion star before it forms the lighter WD, but it is less clear how the latter would be spun-up. The difference in the scenario where $j_2=R_2\omega_2^2$ and the lighter WD is rapidly spinning is however small, with $\sim \mathcal{O}(10^3)$ more systems surviving accretion in the case of weak tides as compared to our fiducial model, and does not affect the main findings of the paper.

Next, given the wide uncertainty on the mass retention efficiency of WDs~\citep{Yaron:2005ys,2013A&A...552A..24B,Wolf:2013hba,2017gacv.workE..56K,Hillman:2020prc}, we revisit the hypothesis that WDs can accrete up to Eddington rate, and limit the maximum accretion rate of WDs to 0.1 and 0.01 of its Eddington rate. In the first case, we obtain results practically identical to the fiducial scenario. In the second case, the number of systems that survive accretion is slightly higher. This comes because the amount of mass and angular momentum lost by the systems increases, helping stabilise the evolution. The global picture remains the same: once the donor overfills its Roche lobe, a violent mass-transfer episode takes place, causing the binary either to merge or to be pushed away, the mass-transfer rate decreases dramatically, and the evolution proceeds in the equilibrium regime discussed in Sec.~\ref{sec:exs}. The stacking of systems around a few mHz is a consequence of the violent episode at the onset of mass transfer and the slow subsequent evolution, and is therefore a robust prediction independent of the details of the model, provided DWDs can survive the violent mass-transfer episode. 

Indeed, it has been suggested that all mass-transferring DWDs might merge \citep{2015ApJ...805L...6S,Pakmor:2022lwn}. To consider the impact of this hypothesis, we remove all accreting systems from our catalogue, and evaluate the stochastic signal and the resolvable binaries for this population. In this scenario, we find no strong dependency on the efficiency of tidal synchronization. This is because, once we remove all accreting systems, without distinguishing between those predicted to be stable and those predicted to be unstable, the remaining population is very similar for all values of $\tau_{s,{\rm ref}}$. For high-spin systems, the stochastic signal we obtain is similar to the one shown in Fig.~\ref{fig:data_analysis} for the 'high-spins, strong tides' case, and the number of resolvable systems is approximately the same as the number of resolvable non-accreting systems in the 'high-spins, strong tides' case, given by the non-hatched part of the histogram. We stress that this latter result is not as trivial as it might seem at first glance. When changing the population, the noise due to the stochastic background changes, changing the criterion within our algorithm to decide if a source is resolvable or not. For the two other scenarios, the number of resolvable systems is similar to the non-accreting portion of the 'low-spins, weak tides' and the stochastic signal is very similar to the one in that scenario, with a slight decrease around a few mHz, due to the absence of accreting systems.

\section{Conclusions}\label{sec:ccl}

In order to prepare for the analysis and astrophysical interpretation of LISA data, it is crucial to bracket the astrophysical predictions for the population of DWDs LISA will observe. In this study, we investigated the influence of tidal torques and mass transfer on the DWD population and their implications for the confusion noise DWDs generate and the number of resolvable binaries.

Starting from a mock catalogue of Galactic DWDs at their formation, predicted by a population synthesis code, we evolved these binaries to the present day using a semi-analytical model that includes the leading order effect of GW radiation, but also tidal torques and mass transfer within the binaries, as well as angular momentum loss due to non-conservation of mass when the accretion rate becomes too large. For binaries undergoing mass transfer, we applied the same simplified criterion found in the literature to determine their stability, which is based on a limit value for the rate of mass lost by the donor. Systems failing this criterion were assumed to merge rapidly and are removed from the catalogue, while those passing it are observed to outspiral and accumulate around a few mHz.

In this model, tidal torques depend on the spins of the WDs and on the tidal synchronization timescale. We define the value of the latter for a reference system at the moment it overfills its Roche lobe, and employ scaling relations based on the mass ratio and separation of the binary to compute it for all systems at any time, facilitating a universal representation of tidal torques. We performed simulations using ten different values of this parameter and explored the uncertainty regarding the spin of WDs at birth by considering three scenarios: one where WDs are born with zero-spin, one where they are already synchronised with the orbit, and one where they are rapidly spinning. We find that the first two cases yield similar results. After generating the mock population, we split the contribution of GWs emitted by DWDs in LISA into a stochastic signal, characterised by its power spectral density, and resolvable binaries.   

 Our findings indicate that weak tidal torques (long timescales) have minimal impact on the population, and incorporating tidal torques and mass transfer into the evolutionary model results in only marginal adjustments compared to scenarios driven solely by GW radiation. Conversely, strong tidal torques can stabilise the evolution and prevent rapid mergers. For WDs born with low spins, we observe a significant accumulation of binaries around a few mHz, which substantially impacts the confusion noise in that frequency region compared to scenarios with GWs alone. Additionally, there is a higher number of resolvable systems in this case. Conversely, for WDs born with large spins, tidal torques tend to slow down the evolution and can even overcome GW radiation, causing the binaries to outspiral even before mass transfer begins. Consequently, the confusion noise due to DWDs becomes almost negligible and very few binaries are resolvable. It is important to note that this latter scenario is likely not representative of the entire population, and predictions for LISA based on it should be interpreted cautiously. On the other hand, observing binaries at very low frequencies ($\lesssim$ 0.1 mHz) with negative frequency derivatives could indicate the presence of a subpopulation of DWDs born with large spins and strong tidal torques. However, due to the low S/N of low-frequency DWDs, we estimate that LISA may not have the sensitivity to detect such a subpopulation, leaving this possibility for future detectors. In a nutshell, strong effects (i.e. strong tides and/or high spins) yield observable signatures. 

 Using the Fisher matrix approximation, we estimated the error on the GW-frequency derivative for resolvable systems. Although the latter is limited, in principle, to the high S/N regime~\citep{Vallisneri:2007ev}, it allows us to estimate the parameters errors of each binary without having to perform a full trans-dimensional Bayesian analysis. We find $\sim 5000$ and $\sim 1000$ accreting systems to have measurable negative GW-frequency derivative in the case of strong and weak tides, respectively. By combining with assumptions on the evolution of DWDs, it should be possible to measure the mass of such systems~\citep{Breivik:2017jip,Yi:2023osk}. Regarding non-accreting systems, we find $\sim 5000$ and $\sim 1000$ of them to have measurable positive GW-frequency derivative in the case of low and high natal spins, respectively. However, for all of them, the systematic error resulting from neglecting astrophysical effects, due to tidal torques in our model, is larger than the statistical error. This renders them unsuitable for chirp mass measurements and tests of GR based on the joint measurement of the gravitational wave frequency and its first derivative in DWDs, unless astrophysical information is incorporated into these measurements, for example by means of universal relations.

We find our results to be robust to different assumptions regarding the evolution of DWDs. The main uncertainty resides in the stability of mass-transferring binaries.  in our study, systems were labelled as unstable under mass transfer when the rate of mass lost by the donor exceeded $10^{-2}M_{\odot}.{\rm yr}^{-1}$. We expect our results to change monotonically with this mass-transfer rate limit: a larger limit rate would lead to more stable accreting binaries and vice versa. We encountered numerical challenges in solving for the evolution within this high-rate regime even within our semi-analytical model, suggesting this is indeed an uncertain regime. However, given the importance of mass transfer on the population of DWDs, a deeper understanding of the stability conditions is essential, in the lines of~\cite{Chen:2022xhn}. The lack of systems at high frequencies ($f_{{\rm GW},f}>20 {\rm mHz}$) is also a consequence of our simplistic treatment: we removed all systems that exceed the limit mass-transfer rate from the catalogue. A better categorisation and treatment of unstable systems could modify this feature. Additionally, for the high-frequency DWDs, which are the ones with measurable frequency derivative, non-linearities in tidal torques should be accounted for~\citep{2020MNRAS.496.5482Y}. Moreover, we restricted ourselves to a single population of DWDs at formation predicted by our population synthesis code. However, there are also numerous uncertainties regarding star formation rate, stellar evolution and binary interactions that would affect the observed population of DWDs. Finally, we did not account for differences in the composition of WDs. It would be interesting to assess how modifications to the mass-radius relation \citep{1961ApJ...134..683H,Suh:1999tg,Panei:1999ji,2024arXiv240313888K} would impact the observed population, creating perhaps distinctive signatures that would allow identifying the type of WDs within a binary. We leave these endeavours for future research. 

Our results complement the findings of~\cite{Scaringi:2023xpm}, where the authors found that cataclysmic variables, formed of a WD accreting from a main-sequence or sub-stellar star, lead to measurable deviations in the confusion noise below 1 mHz. Together, these studies highlight the need for flexible tools in LISA data analysis, in order to account for all possibilities in the widely unknown population of compact objects in the Galaxy. Our exploration provides insights into potential parametrisations of the DWD population for population analysis with LISA. We suggest representing the population of resolvable DWDs as a power-law contribution in frequency with positive $\dot{f}_{{\rm GW},f}$ and a Gaussian around a few mHz with negative $\dot{f}_{{\rm GW},f}$. For the stochastic signal, a flexible function accommodating the various possibilities demonstrated in our study would be suitable, possibly employing the linear interpolation scheme proposed with variable number of knots proposed in~\cite{Toubiana:2023egi}. In order to gain informative power, it would be valuable to establish relationships between the properties of resolvable and non-resolvable populations, a task we defer to future investigations.


\section*{Acknowledgements}

We are thankful to V. Korol for her valuable suggestions during the preparation of this work. We are also thankful to the efforts of the LDC Working Group for the software availability and help. N.Karnesis acknowledges the funding from the European Union’s Horizon 2020
research and innovation programme under the Marie Skłodowska-Curie grant agreement No 101065596. Astrid Lamberts is supported by the
ANR COSMERGE project, grant ANR-20-CE31-001 of the French Agence Nationale de la Recherche. This
work was supported by the ‘Programme National des Hautes Energies’ (PNHE)
of CNRS/INSU co-funded by CEA and CNES.

 \bibliographystyle{aa} 
\bibliography{Ref} 

\appendix

\section{Oscillations in the evolution}\label{app:res}

\begin{figure}[h!]
\centering
 \includegraphics[width=0.49\textwidth]{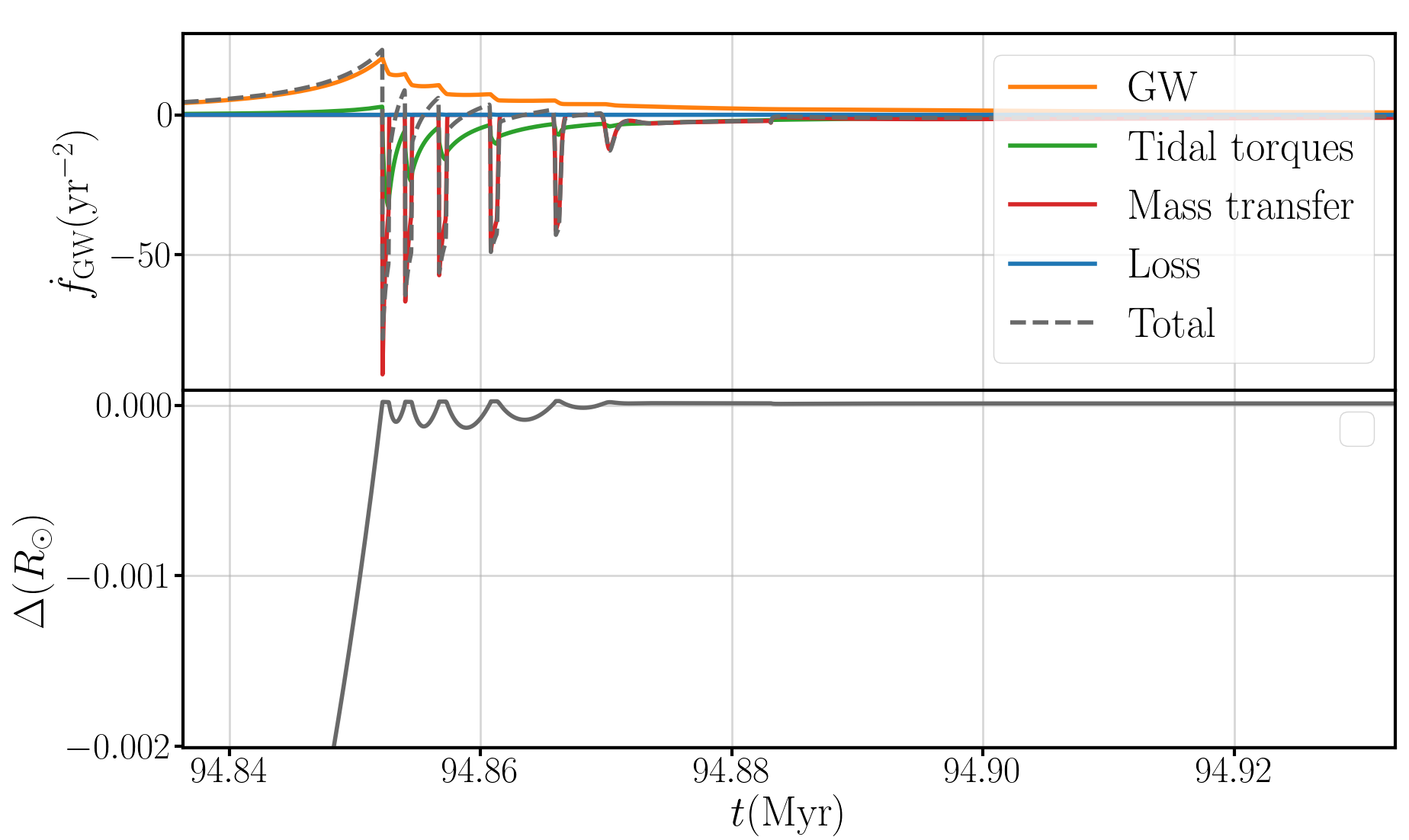}\\
 \centering
 \caption{Same as Fig.~\ref{fig:ex_fdot}, but with $\tau_{s,{\rm ref}}=10^{2}\ {\rm yr}$, and we show in parallel the evolution of the Roche lobe overflow.}\label{fig:ex_fdot_delta}
\end{figure}

For certain combinations of WD masses and tidal synchronization timescale, we observe very pronounced oscillations in the binary separation due to the donor successively overfilling and 'underfilling' its Roche lobe. A similar behaviour is discussed in~\cite{Marsh:2003rd,Gokhale:2006yn,2015ApJ...806...76K}. This oscillatory behaviour is illustrated in Fig.~\ref{fig:ex_fdot_delta}, where we show the evolution of $\dot{f}_{{\rm GW}}$ for the same system as in Fig.~\ref{fig:ex_fdot}, but taking a shorter tidal synchronization timescale, $\tau_{s,{\rm ref}}=10^{2}\ {\rm yr}$. The explanation behind these oscillations is as follows. After the donor first overfills its Roche lobe and mass transfer starts, the distance between WDs increases, at a larger rate than in the previous case, due to the larger negative (positive for the separation) contribution of tidal terms (see upper panel of Fig.~\ref{fig:ex_fdot_delta}), the Roche lobe expands faster than the donor, which no longer overfills its Roche lobe. GWs take over and bring the system closer again, until the donor overfills its Roche lobe and mass transfer resumes. This succession of events keeps on repeating until the donor is light enough and its radius expands faster than the Roche lobe, so that mass transfer is not interrupted. In total, this oscillatory stage would last a few tens of Gyr, alternating oustpiralling regimes of $\sim$ 1 Gyr and inspiralling regimes of $\sim $ 5 Gyr.

\begin{figure}[h!]
\centering
 \includegraphics[width=0.49\textwidth]{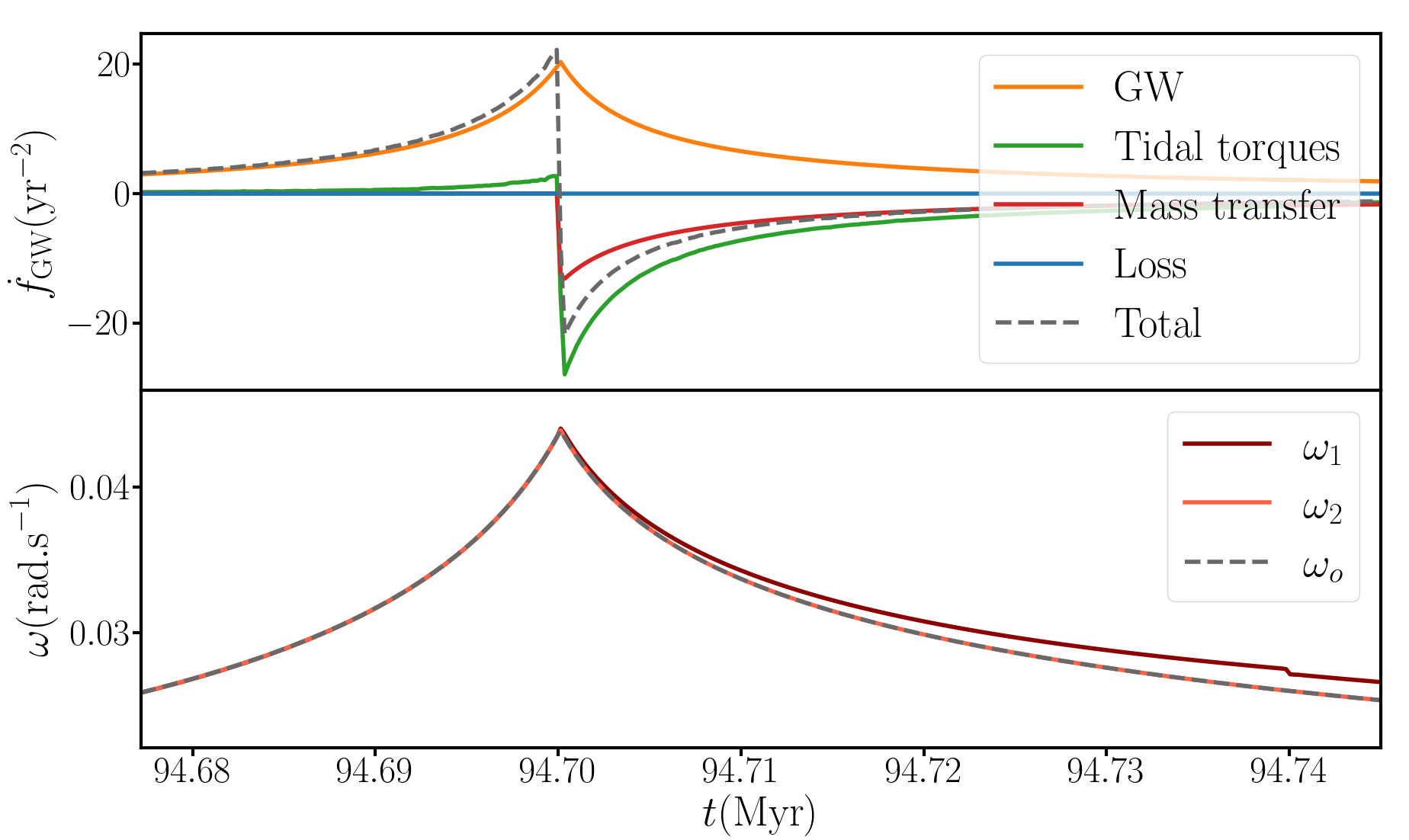}\\
 \centering
 \caption{Same as Fig~\ref{fig:ex_fdot}, but with $\tau_{s,{\rm ref}}=1\ {\rm yr}$, and we show in parallel the evolution of the accretor and the donor spins, as well as the orbital frequency.}\label{fig:ex_fdot_omega}
\end{figure}

If we consider an even shorter $\tau_{s,{\rm ref}}$, we no longer observe these oscillations. The reason for this is as follows. Before the onset of mass transfer, tidal torques are very efficient at synchronising the WDs with the orbit, as can be seen in Fig.~\ref{fig:ex_fdot_omega}. Then, mass transfer starts, leading to an additional torque (first term in Eq.~\eqref{eq:omega}), which desynchronises the WDs (in this example, mostly the accretor). The tidal torque then becomes very strong, due to the small synchronization timescale, and drives the binary apart. Once the tidal term becomes sub-dominant, the donor is light enough so that the Roche lobe does not expand faster than its radius, preventing the oscillatory behaviour described above.

\end{document}